\documentclass[english]{elsarticle}
\usepackage{amsmath}
\usepackage{graphicx}

\makeatletter

\usepackage{amsfonts}

\usepackage{amsthm}

\usepackage{mathrsfs}

\usepackage{graphics}

\usepackage{lscape}

\usepackage{changebar}

\usepackage{epsfig}

\newcommand{\etal}{\textit{et al. }}

\renewcommand{\vec}[1]{\mathbf{#1}}

\newcommand{\chib}{\mbox{\boldmath$\chi$}}
\newcommand{\xib}{\mbox{\boldmath$\xi$}}

\newcommand{\deriv}[2]{\frac{\partial #1}{\partial #2}}

\makeatother

\usepackage{babel}

\begin{document}

\title{An Eulerian approach to the simulation of deformable solids: Application to finite-strain elasticity}



\author[Harvard]{Ken Kamrin\corref{cor1}}
\ead{kkamrin@seas.harvard.edu}
\ead[url]{http://people.seas.harvard.edu/$\sim $kkamrin/}
\author[MIT]{Jean-Christophe Nave}
\ead{jcnave@mit.edu}
\ead[url]{http://www.mit.edu/$\sim $jcnave/}

\cortext[cor1]{Corresponding author.}

\address[Harvard]{School of Engineering and Applied Sciences, Harvard University Cambridge, MA 01238, USA}
\address[MIT]{Department of Mathematics, Massachusetts Institute of Technology,
Cambridge, MA 02139, USA}
\begin{abstract}
  We develop a computational method based on an Eulerian field
  called the {}``reference map'', which relates the current location
  of a material point to its initial.  The reference map can be
  discretized to permit finite-difference simulation of large
  solid-like deformations, in any dimension, of the type that might
  otherwise require the finite element method. The generality of the
  method stems from its ability to easily compute kinematic quantities
  essential to solid mechanics, that are elusive outside of Lagrangian
  frame. This introductory work focuses on large-strain, hyperelastic
  materials. After a brief review of hyperelasticity, a
  discretization of the method is presented and some non-trivial
  elastic deformations are simulated, both static and dynamic.  The
  method's accuracy is directly verified against known analytical
  solutions.
\end{abstract}


\maketitle

\section{Introduction}

Two perspectives are used for modeling solid and amorphous materials
in the continuum limit. Physicists and rheologists tend to prefer an
Eulerian treatment where the material can be viewed essentially as a
fluid equipped with state variables that provide ``memory'' (see
\cite{falk98,lemaitre02}).  On the other hand, many in the engineering
community prefer a Lagrangian description, where the material body is
decomposed as a network of volume elements that deform under stress
(see \cite{anand05,gurtin81}).

Continuum mechanical laws derive from point-particle mechanics applied
to a continuum element. The primitive form is resultantly Lagrangian,
though an Eulerian conversion can always be asserted--- one rewrites
the constitutive laws in rate form and expands all material time
derivatives in terms of fixed-space derivatives. Be that as it may, a
rigorous Eulerian switch can be a painstaking mathematical task.
This is especially true of solid-like constitutive laws, which often
depend on nonlinear tensor operations and coupled history-dependent
state variables, leading to unduly complicated Eulerian rate expansions
\cite{hoger86}.

To dodge these difficulties, those preferring Eulerian-frame have
generally resorted to approximations or added conditions that simplify
the final constitutive form. While sometimes warranted, the connection
back to Lagrangian mechanics becomes clouded, complicating the process
of deriving physically motivated constitutive behavior.

In this paper, a field we call the ``reference map'' is utilized to
construct and implement solid-like constitutive laws in Eulerian-frame
with \emph{no} added approximations. The way the map provides
``memory'' to the system admits immediate computation of kinematic
variables crucial to Lagrangian solid mechanics. To maintain a clear
presentation, several avenues of motivation are first provided that
discuss the necessary laws of continuum mechanics and the basic
quantities of solid kinematics. The theory of large-strain elasticity,
hyperelasticity, is then sketched. In particular, by enabling quick
access to the \emph{deformation gradient} tensor, the reference map
can be used to accurately compute solid deformations without the
approximations, ambiguities, or pre-conditions of other Eulerian
approaches. Three non-trivial deformations are then simulated to
verify these points.

\section{Basics}\label{basics}

In Eulerian frame, the flow or deformation of a canonical continuous material
can be calculated by solving a system of equations that includes:
\begin{align}
  & \rho_{t}+\nabla\cdot(\rho\vec{v})=0\label{basic1}\\
  & \nabla\cdot\vec{T}+\rho\vec{g}=\rho(\vec{v}_{t}+\vec{v}\cdot\nabla\vec{v})\label{basic2}\\
  & \vec{T}^{T}=\vec{T}.\label{basic3}\end{align} The first equation
upholds mass conservation, and the next two, respectively, uphold
conservation of linear and angular momentum. The flow is described by
the velocity field $\vec{v}(\vec{x},t)$ and the stresses by the Cauchy
stress tensor $\vec{T}(\vec{x},t)$, which includes pressure
contributions. A consitutive law is then asserted to close the system
of equations.

\section{Solid kinematics}\label{solid}

We ultimately intend our approach to apply to any material with a
``solid-like'' constitutive law. By solid-like, we mean specifically
laws that express the stress tensor in terms of some kinematic
quantity that measures the local deformation from some nearest relaxed
state. This trait reflects the microscopic basis of solid stress as
arising from potential energy interactions between material
microconstituents. The simplest solid-like response is isothermal
elasticity, where total deflection under loading immediately
determines the stresses within \cite{gurtin81}.  A less basic example
would be elasto-plasticity, where internal stresses derive from a
small elastic component of the total strain. Here, the nearest relaxed
state can differ from the original unstressed state and may depend on
evolving state parameters, temperature, and/or rate \cite{lee69}.

To encompass the broad definition above, a continuum description for
solid-like materials necessitates a rigorous way of tracking local
relative displacements over some finite time period. Without making
any ``small displacement'' approximations, a general and robust
continuum framework calls for a kinematic field $\chib$ known as the
\emph{motion function}. Suppose at time $t=0$, that a body of material
is in an unstressed \emph{reference configuration}. The body then
undergoes a deformation process such that at time $t$, an element of
material originally at $\vec{X}$ has been moved to $\vec{x}$.  The
motion is defined by $\vec{x}=\chib(\vec{X},t)$. We say that the body
at time $t$ is in a \emph{deformed configuration}.

The motion can be used to define the \emph{deformation gradient}
$\vec{F}$, which is of crucial importance in continuum solid
mechanics: \begin{equation}
  \vec{F}(\vec{X},t)=\deriv{\chib(\vec{X},t)}{\vec{X}}\ \ \ \text{or}\
  \ \ F_{ij}=\deriv{\chi_{i}(\vec{X},t)}{X_{j}}.\end{equation} Note
that we use $\nabla$ for gradients in $\vec{x}$ only, and always write
gradients in $\vec{X}$ in derivative form as above.  As per the chain
rule, the $\vec{F}$ tensor describes local deformation in the
following sense: If $d\vec{X}$ represents some oriented, small
material filament in the reference body, then the deformation process
stretches and rotates the filament to $d\vec{x}=\vec{F}\ d\vec{X}$ in
the deformed body. Also, the evolution of $\vec{F}$ can be connected
back to the velocity gradient via \begin{equation}
  \dot{\vec{F}}=\left(\nabla\vec{v}\right)\vec{F}\label{F_evolution}\end{equation}
where we use $\dot{}$ for material time derivatives.

Since $\text{det}\vec{F}>0$ for any physical deformation, the
deformation gradient admits a polar decomposition $\vec{F}=\vec{RU}$
where $\vec{R}$ is a rotation, and $\vec{U}$ is a symmetric positive
definite ``stretch tensor'' obeying
$\vec{U}^2=\vec{F}^{T}\vec{F}\equiv\vec{C}$.

\section{General finite-strain elasticity}\label{finite} 
To demonstrate the use and simplicity of the method, this paper shall
focus on one broad class of materials: large-strain, 3D, purely
elastic solids at constant temperature. A thermodynamically valid
constitutive form for such materials is derivable with only minimal
starting assumptions.  Known as \emph{hyperelasticity} theory, it has
become the preeminent elasticity formulation in terms of physicality
and robustness.  Though other elasticity formulations exist (e.g.
hypoelasticity and other stress-rate models) the next section will
recall how these are in fact specific limitting approximations to
hyperelasticity theory. A brief review of hyperelasticity is provided
below to establish the key results and demonstrate the physical basis
of the theory (see \cite{gurtin81} for details).
An analysis of more complex solid-like behaviors
(e.g. elasto-plasticity, hardening, thermal elasticity) is left as
future work.

In essence, one seeks a noncommittal 3D extension of 1D spring mechanics,
where total relative length change determines the force in a fashion
independent of deformation path. To institute this, presume that
the Helmholtz free-energy per unit (undeformed) volume $\psi$ and
Cauchy stress $\vec{T}$ both depend only on the local deformation:
\begin{equation}
\psi=\hat{\psi}\left(\vec{F}\right)\ ,\ \vec{T}=\hat{\vec{T}}\left(\vec{F}\right)\label{general}\end{equation}
 where $\hat{}$ is used to designate constitutive dependences on
kinematic quantities. We also assume that if no deformation has occurred
(i.e. $\vec{F}=\vec{1}$), then $\vec{T}=\vec{0}$.

Some helpful physical principles refine these dependences
immensely. We enforce \emph{frame-indifference} by restricting the
dependences to account for rotations. Suppose a material element is
deformed by some amount, and then the deformed element is rotated. By
the frame-indifference principle, the rotation should not affect the
free-energy, and should only cause the stress to co-rotate. This
ultimately restricts Eqs \ref{general} to \begin{equation}
  \psi=\hat{\psi}(\vec{C})\ \ \text{and}\ \ \vec{T}=\vec{R}\
  \hat{\vec{T}}(\vec{C})\
  \vec{R}^{T}.\label{less_general}\end{equation}

Next, we enforce \emph{non-violation of the second law}. A
continuum-level expression of the isothermal second law of
thermodynamics can be written as the dissipation law \begin{equation}
  \rho\dot{\psi}-\vec{T}:\vec{D}\le0\label{second}\end{equation} where
$\vec{D}=(\nabla\vec{v}+(\nabla\vec{v})^{T})/2$ is the deformation
rate, familiar from fluid mechanics. Following a procedure originally
developed by Coleman and Noll \cite{coleman63}, one can prove
mathematically that Eqs \ref{less_general} uphold Ineq \ref{second}
under all imposable deformations only if: \begin{equation}
  \vec{T}=2(\text{det}\vec{F})^{-1}\
  \vec{F}\deriv{\hat{\psi}(\vec{C})}{\vec{C}}\vec{F}^{T}.\label{elasticity}\end{equation}
Likewise, $\vec{C}=\vec{0}$ must correspond to a local minimum of
$\psi$. Eq \ref{elasticity} along with the zero deformation hypotheses
compose the theory of hyperelasticity.

The above argument demonstrates how the assertions of
frame-indifference and the second-law require that the assumed
dependences of Eq \ref{general} refine to the form of Eq
\ref{elasticity}. Each valid choice of $\hat{\psi}$ gives an
elasticity law that could represent a continuous elastic solid. The
``deductive approach'' above has become a frequently used tool in
materials theory.

\section{Some past Eulerian attempts}\label{past}

To use hyperelasticity, or deductive solid modeling in general, the
ability to calculate $\vec{F}$ during a deformation is crucial. In
Lagrangian-frame, each point is ``tagged'' by its start point $\vec{X}$, so
$\vec{F}$ can always be computed by differentiating current location
against initial. In Eulerian the problem is more subtle, as knowledge
of past material locations must somehow be procured. As suggested in
\cite{liu01}, $\vec{F}$ can be directly evolved by expanding the
material time derivative in Eq \ref{F_evolution},
giving \begin{equation}
  \vec{F}_{t}+\vec{v}\cdot\nabla\vec{F}=(\nabla\vec{v})\vec{F}.\label{F_euler}\end{equation}
Unfortunately, this cannot be used to solve the general boundary value
problem. The term $\nabla \vec{F}$ can only be computed adjacent to
boundaries if $\vec{F}$ is prescribed as a boundary condition. To
assign $\vec{F}$ at a boundary implies that the derivative of motion
in the direction orthogonal to the surface can be controlled. In the
general boundary value problem, this information is outside the realm
of applicability; stress tractions and displacements/velocity
conditions can be applied at boundaries, but how these quantities
change orthogonal to the surface arises as part of the deformation
solution.

Another approach that also advects the $\vec{F}$ tensor directly (more
factually the tensor $\vec{F}^{-1}$) is the Eulerian Godunov method of
Miller and Colella \cite{miller01}.  The method solves for elastic or
elasto-plastic solid deformation by treating the equations as a system
of conservation laws with a nonconservative form for the advection of
$\vec{F}^{-1}$.  It is a sophisticated, high-order method and has had
success representing solid dynamics and deformation, but is aimed
primarily at unbounded domains where implementation of a
boundary condition on $\vec{F}$ is unneeded.

For pure elasticity, several Eulerian, rate-based approaches have
been developed that avoid directly referring to $\vec{F}$ but add
in several approximations/assumptions. Begin by presuming isotropy.
It can be shown that this reduces Eq \ref{elasticity} to \begin{equation}
\vec{T}=\frac{2}{\sqrt{I_{3}}}\left[I_{3}\deriv{\hat{\psi}}{I_{3}}\vec{1}+\left(\deriv{\hat{\psi}}{I_{1}}+I_{1}\deriv{\hat{\psi}}{I_{2}}\right)\vec{B}-\deriv{\hat{\psi}}{I_{2}}\vec{B}^{2}\right]\label{elasticity_final}\end{equation}
 where $I_{1},I_{2},I_{3}$ are the principal invariants of the left
Cauchy-Green tensor $\vec{B}=\vec{F}\vec{F}^{T}$ \cite{gurtin81}.

One way to uphold Eq \ref{elasticity_final} involves first defining a
\emph{strain measure} $\vec{E}=\vec{E}(\vec{B})$ that, among other
features, must asymptote to $\vec{E}=0.5(\vec{F}+\vec{F}^{T})-\vec{1}$
in the small displacement limit $|\vec{F}-\vec{1}|\ll1$. To linear
order in $\vec{E}$, the elasticity law can then be written as 
\begin{equation}\vec{T}=2G\vec{E}+\lambda(\text{tr}\vec{E})\vec{1}\equiv\mathscr{C}(\vec{E}).\end{equation} 
Taking the material time derivative gives
$\dot{\vec{T}}=\mathscr{C}(\dot{\vec{E}})$. The chain rule on
$\dot{\vec{E}}$ generally leads to a long expression in terms of
$\vec{F}$ and $\dot{\vec{F}}$, which can ultimately be rewritten as
some function of $\vec{T}$ and $\nabla\vec{v}$.

Eulerian expansion of $\dot{\vec{T}}$ introduces a term
$\vec{v}\cdot\nabla\vec{T}$. Once again, the same problem as that
encountered in Eq \ref{F_euler} occurs; to compute
$\nabla\vec{T}$, the full Cauchy stress tensor $\vec{T}$ must be
assigned at the boundary.  While certain components of $\vec{T}$ can
be controlled at a boundary--- namely the traction vector
$\vec{T}\hat{\vec{n}}_{\text{bound}}$ --- the components describing
stresses along a plane orthogonal to the surface cannot, in general,
be prescribed.

To dodge this difficulty the term $\vec{v}\cdot\nabla\vec{T}$ is
presumed to be negligible. As a consequence of neglecting stress
convection, one accepts certain errors in representing dynamic
phenomena. Ultimately, what remains is an Eulerian constitutive
relation for the evolution of $\vec{T}$ \begin{equation}
  \vec{T}_{t}=\mathscr{C}\left(A(\vec{T},\nabla\vec{v})\right)\label{rate2}\end{equation}
where the function $A$ derives from the choice of strain-measure. While
$\vec{D}$ is sometimes called the ``strain-rate'', we note that it is
not the time rate of change of a valid strain measure; the axes of
$\int\vec{D}\ dt$ do not rotate with the material. However,
$\dot{\vec{E}}\approx\vec{D}$ in the small displacement limit for all
strain definitions. Assuming small displacement and small rate of
volume change, Eq \ref{rate2} reduces to a simple form known as
\emph{hypoelasticity} \begin{equation}
  \vec{T}^{\circ}=\mathscr{C}(\vec{D})\label{hypo}\end{equation} where
$\vec{T}^{\circ}$ is an ``objective stress rate'' equal to
$\vec{T}_{t}$ plus extra terms that depend on the choice of strain
measure.

Hypoelasticity can be seen as a specific approximation to a physically
derived isotropic hyperelasticity law. Be that as it may, Eq
\ref{hypo} is oftentimes asserted as a starting principle by assigning
$\vec{T}^{\circ}$, sometimes arbitrarily, from a list of commonly used
stress rates--- e.g. Jaumann rate, Truesdell rate, Green-Naghdi rate
(see \cite{meyers06} for a detailed review) ---thereby cutting off the
connection to hyperelasticity. In fact, there are infinitely many
stress rate expressions upholding frame-indifference that qualify as
objective hypoelastic rates.

Rate forms for elasticity require the assumptions and approximations
listed herein, which limit their applicability. The neglect of stress
convection can pay heavy consequences when attempting to represent
waves or other dynamic phenomena. While Eq \ref{rate2} is fairly
general for isotropic linearly elastic materials, the resulting
equations usually require tedious calculation that must be redone if
the stress/strain relation is changed.  Even in the small strain
limit, hypoelasticity's presumptions of linearity and isotropy poorly
represent some common materials. For instance, granular matter is
nonlinear near zero strain (due to lack of tensile support), and
crystalline solids are not isotropic. Rate elasticity, if used as a
first principle, also offers no physical basis to account for
thermodynamics, making it troublesome for theories of thermalized or
non-equilibrium materials.

\section{The reference map}
To sidestep these issues, we now describe a new Eulerian approach to
solid mechanics. The key is to utilize a fixed-grid field that admits
a direct computation of $\vec{F}$. Define a vector field called the
\emph{reference map} $\xib(\vec{x},t)$ by the evolution
law: \begin{equation} \xib_{t}+\vec{v}\cdot\nabla\xib=\vec{0}\ \
  \text{for}\ \
  \xib(\vec{x},t=0)=\vec{x}=\vec{X}.\label{advect}\end{equation} This
advection law implies that $\xib$ never changes for a tracer moving
with the flow. Combined with the initial condition, the vector
$\xib(\vec{x},t)$ indicates where the material occupying $\vec{x}$ at
time $t$ originally started.

By the chain rule, a material filament obeys
$d\vec{X}=(\nabla\xib)d\vec{x}$. Thus, \begin{equation}
  \vec{F}=\left(\nabla\xib\right)^{-1}.\label{Fnew}\end{equation}
Altogether, Eqs \ref{basic2}, \ref{elasticity}, \ref{advect}, and
\ref{Fnew}, along with the kinematic expression for the density
$\rho=\rho_{0}(\text{det}\vec{F})^{-1}$, compose an Eulerian system
that solves exactly for hyperelastic deformation.

In essence, what we are suggesting is to obtain solid stress in a
fashion similar to fluid stress. For fluids, the (shear) stress is
given by $\vec{D}$, which is computed from the gradient of $\vec{v}$.
Here, we advect the primitive quantity $\xib$ and use its gradient to
construct $\vec{F}$. The stress is then obtained from $\vec{F}$ by the
constitutive law.  This approach alleviates many of the complications
discussed previously that arise when attempting to directly advect a
tensorial quantity like $\vec{F}$ or $\vec{T}$.

In particular, unlike the advection law of Eq \ref{F_euler}, the
reference map is easily definable on boundaries provided complete
velocity/displacement boundary conditions. That is, if a boundary
point originally at $\vec{X}_{b}$ is prescribed a displacement
bringing it to $\vec{x}_{b}$ at time $t$, then
$\xib(\vec{x}_{b},t)=\vec{X}_{b}$.  We also note that $\xib$ is an
integral quantity of $\vec{F}$ and thus a smoother function. We expect
this property to be of benefit numerically compared to methods that
directly advect $\vec{F}$ or $\vec{T}$.

The notion of a map that records initial locations of material has
been defined by others in various different contexts.  To these
authors' knowledge, it has never been used for the purposes of solving
 solid deformation as described above. Koopman \etal
\cite{koopman08} use an ``original coordinate'' function akin to our
reference map in defining a pseudo-concentration method for flow
fronts.  The inverse of $\chib$ at time $t$, which is indeed
equivalent to the $\xib$ field, is also discussed in Belytschko
\cite{belytschko00} for use in finite element analysis.

\section{Implementation}\label{implementation}

In this section, we describe the discretization of the above system of
equations. Our general strategy is to first evaluate $\vec{T}$, then
update $\vec{v}$ using Eq \ref{basic2}, and finally evolve $\xib$ with
\ref{advect}. Time derivatives in Eqs \ref{basic2}, and \ref{advect}
are discretized as a simple Euler step,
\[
\partial_{t}\xib(\vec{x},t)=\left(\xib(\vec{x},t+\Delta
  t)-\xib(\vec{x},t)\right)/\Delta t.\] On a two-dimensional grid, with grid
spacing $h$, the velocity $\vec{v}$ and reference map $\xib$ are
located at corner points $(i,j)$, while stresses $\vec{T}$ are located
at cell centers, $(i+\frac{1}{2},j+\frac{1}{2})$.  Thus, away from any
boundary, we can compute by finite difference $\partial_{x}\xib$ at
the mid-point of horizontal grid edges, and similarly,
$\partial_{y}\xib$ on vertical grid edges,
\begin{equation}
\partial_{x}\xib_{(i+\frac{1}{2},j)}=\left(\xib_{(i+1,j)}-\xib_{(i,j)}\right)/h.\label{eq:dxi_dx_discretization}
\end{equation}
We obtain $\nabla\xib$ at cell centers by averaging, allowing us
to compute the deformation gradient tensor $\vec{F}$ using Eq \ref{Fnew},
and thus $\vec{B}$. We now can define stresses at cell centers by
specifying the hyperelasticity law. We compute $\partial_{x}\vec{T}$
at the mid-point of vertical grid edges, and similarly, $\partial_{y}\vec{T}$
on horizontal grid edges, \[
\partial_{x}\vec{T}_{(i,j+\frac{1}{2})}=\left(\vec{T}_{(i+\frac{1}{2},j+\frac{1}{2})}-\vec{T}_{(i-\frac{1}{2},j+\frac{1}{2})}\right)/h.\]
As a result, we obtain $\nabla\cdot\vec{T}$ at cell corners, where
$\vec{v}$ is stored. Finally, $\nabla\vec{v}$ in equation
\ref{basic2}, can be discretized in the same manner as
$\nabla\xib$. Additionally, since Eq \ref{advect} is an advection
equation, we use a WENO discretization for $\nabla\xib$ to guarantee stability
\cite{liu94}.

In order to solve the system with irregular boundaries, we introduce a
level set function, $\phi(x)$. We define $\phi(x)>0$ inside the solid,
$\phi(x)\leq0$ outside, thus implicitly representing the domain
boundary as the zero level set of $\phi(x)$. Choosing $\phi(x)$ to be
a signed distance function, i.e. $\|\nabla \phi(x)\|=1$, allows us to
compute the cut-cell length $\alpha$, with $0\leq\alpha\leq1$.  For
example, if the boundary cuts a horizontal cell edge,
i.e. $\phi_{left}\cdot\phi_{right}\leq 0$, then
$\alpha=\frac{\mid\phi_{left}\mid}{\mid\phi_{left}\mid+\mid\phi_{right}\mid}.$
Here, we must change the discrete derivatives in Eq
\ref{eq:dxi_dx_discretization}. Assuming that
$\phi_{left}=\phi_{(i,j)}>0$, \[
\partial_{x}\xib_{(i+\frac{1}{2},j)}=\left(\xib_{right}-\xib_{(i,j)}\right)/\alpha h\]
where $\xib_{right}$ is a boundary condition for $\xib$. Other derivatives
near boundaries are treated in the same manner.

\section{Results}

\begin{figure}
(a) \epsfig{file=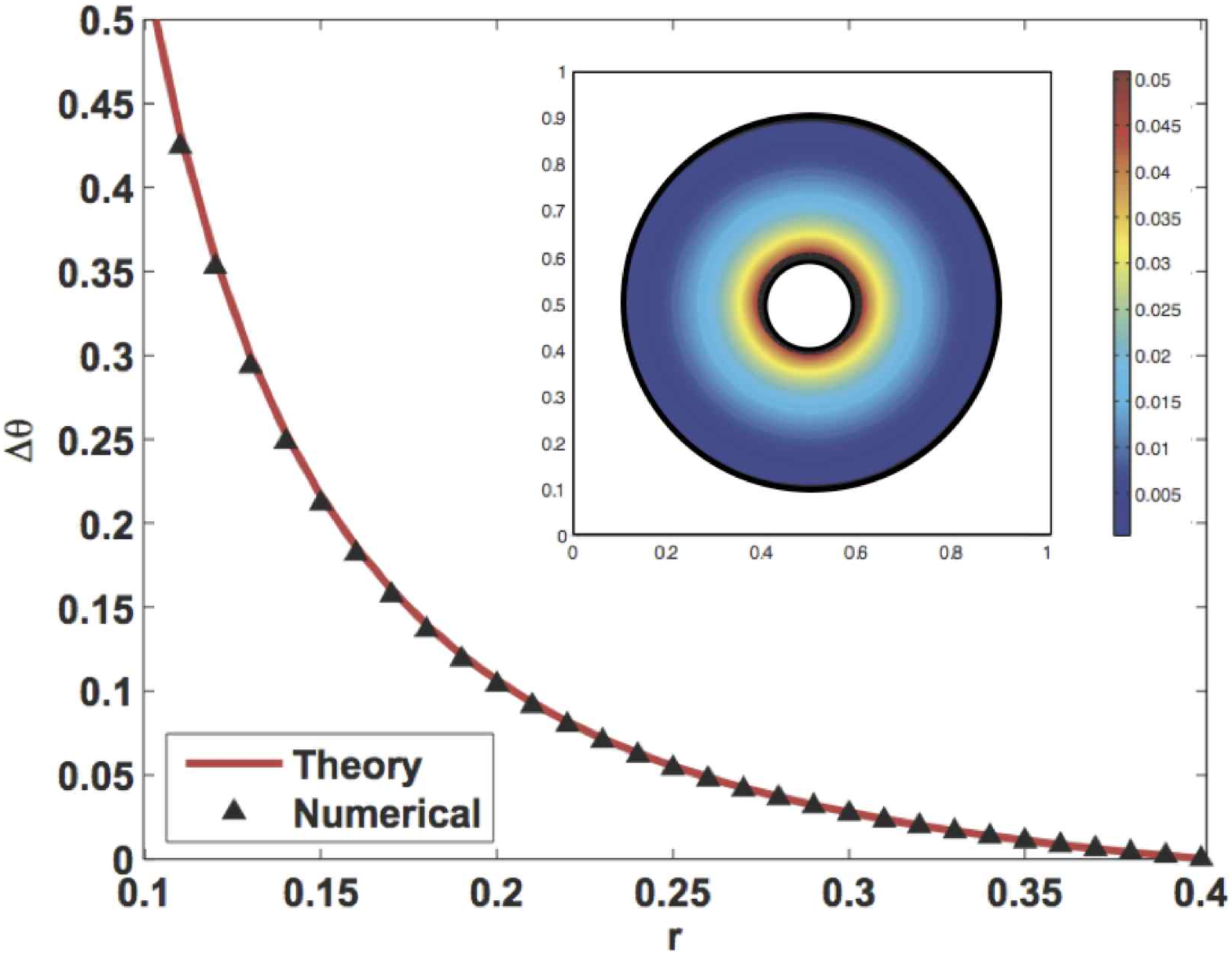, width=2.8in}
(b)\epsfig{file=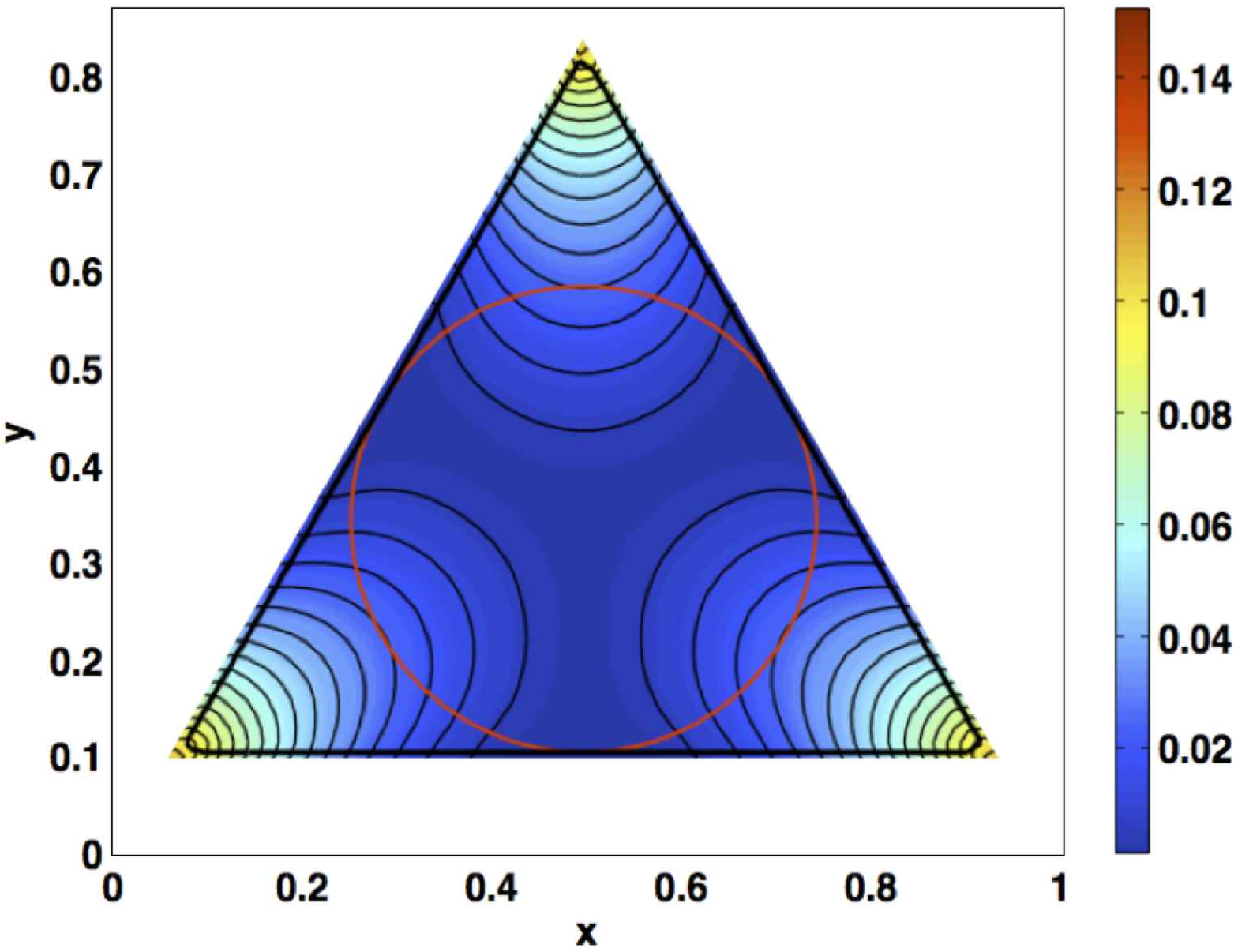, width=2.8in}
\caption{(a) Angular displacement of washer after inner wall
  rotation. Inset displays the scalar displacement field; note high
  radial symmetry though the scheme uses a fixed
  \emph{cartesian} grid. (b) Displacement field as a function of final location after an
  elastic disk (red outline) is stretched into a
  triangle. Both tests were performed on a 100x100 grid.}\label{washer_triangle}
\end{figure}
\subsection{Static solutions}

In this section we present two large-deformation numerical tests, both
in plane-strain for simplicity. To rigorously test the method, each
case models a non-trivial, inhomogeneous deformation. First, we solve
our system in a circular washer geometry for which the outer wall is
fixed and the inner wall is rotated over a large angle.  Under the
Levinson-Burgess hyperelasticity law (see below), this environment has
an analytical solution, which we use to verify the
consistency/correctness of the method. Second, we solve the
deformation of a disk being stretched into a triangular shape, also
utilizing the Levinson-Burgess law. This has no analytical solution,
and demonstrates the method's applicability in cases where the reference
and deformed boundary sets differ. We focus here on static solutions,
obtained by enforcing the boundary conditions and waiting for
transients to pass.  Artificial viscosity was added to the stress law
to expedite collapse to the static solution.

The Levinson-Burgess free-energy function, after application of Eq
\ref{elasticity_final}, induces the following stress law under
plane-strain conditions \cite{haughton93}
\begin{equation}
  \vec{T}=f_{1}(I_{3})\ \vec{B}+f_{2}(I_{1},I_{3})\ \vec{1}\label{eq:Levinson-Burgess}\end{equation}
where $I_{1}=\text{tr}\vec{B}$, $I_{3}=\text{det}\vec{B}$ are invariants
of the $\vec{B}$ tensor, with $f_{1}(I_{3})=G\left(3+1/I_{3}\right)/(4\sqrt{I_3})$ and $f_{2}(I_{1},I_{3})=G\sqrt{I_{3}}\left(\kappa/G-1/6+(1-I_{1})/(4I_{3}^{2})\right)-G/3-\kappa$. In the unstrained state, $\kappa$ and $G$ represent the bulk and shear moduli. Throughout, we use $\kappa=G=$100 kPa.

\paragraph{Circular washer shear} The analytical static displacement
field is $\Delta\theta=A-B/r^{2}$ and $\Delta r=0$ where, $A$ and $B$
are constants that fit the boundary conditions
$\Delta\theta_{in}=\pi/6$ and $\Delta \theta_{out}=0$. The graph in
Figure \ref{washer_triangle}(a) shows excellent agreement between our
numerical solution (sampled along the central horizontal
cross-section) and the analytical.  We have observed equally high
agreement levels when the inner wall rotation angle is varied.

\paragraph{Stretched Disk}

The unstressed material shape is a disk that is inscribed perfectly
within its final equilateral triangular shape. For $\vec{x}_b$ on the
triangle edge, the boundary condition for the final deformed body is
$\xib(\vec{x}_b)=\frac{r_0\vec{x}_b +
  d(\vec{x}_b)\vec{x}_{cen}}{r_{0}+d(\vec{x}_b)}$ where
$\vec{x}_{cen}$ is the disk center, $r_0$ the disk radius, and
$d(\vec{x}_b)$ the distance between the disk edge and the triangle as
measured along the radial segment containing $\vec{x}_b$. Hence, each
point on the disk edge is moved outward radially to the triangle. The
final, static displacement field is shown in figure \ref{washer_triangle}(b).

\subsection{Dynamic solutions}

In this section, we display the method's ability to accurately track
the motion of a large-strain compression wave.  Recall from section
\ref{past}, that dynamic correctness is sacrificed in many stress-rate
models by neglecting stress convection terms.  The reference map on
the other hand, enables high-accuracy representation of
elasto-dynamics as shall now be demonstrated.

In order to check for the ability of the method to handle dynamic
situations, we choose to produce an analytical solution for an elastic
compresssion wave. For this purpose, consider a material obeying the
following large-strain elasticity law,

\[
\vec{T}=\kappa(\vec{V}-\vec{1})\]
 for $\vec{V}\equiv\sqrt{\vec{B}}$ the left stretch tensor. The material
body is a rectangular slab constrained in the thickness direction
(i.e. plane-strain conditions). The unstressed material density is
uniform and has a value $\rho_{0}$.

Under this constitutive law, the following $\xib$ and $\vec{v}$ fields
give an exact, analytical solution for a rightward moving compression
wave passing through the slab: \begin{equation}
  \hat{\vec{x}}\cdot\xib\left(x,y,t\right)=x+\frac{1}{2}\text{Erf}(x-ct)\label{eq:xi_exact_solution}\end{equation}
 \begin{equation}
   \hat{\vec{x}}\cdot\vec{v}\left(x,y,t\right)=c\left(1-\cfrac{1}{1+\frac{1}{\sqrt{\pi}}e^{-(x-ct)^{2}}}\right).\label{eq:v_exact_solution}\end{equation}
 Due to symmetry, the $\hat{\vec{y}}$ and $\hat{\vec{z}}$ components
 of both fields do not change from their initial, unstressed values.
 The constant $c=\sqrt{\kappa/\rho_{0}}$ is the wave speed. This solution
 invokes a large-strain deformation with compressive strain as high
 as $|\hat{\vec{x}}\cdot(\vec{V}-\vec{1})\hat{\vec{x}}|\approx36\%$
 at the center of the pulse. Consequently, this represents a realistic
 test of the ability of the present approach to tackle dynamic effects
 as well as large-strain deformation.

The equations are discretized in space as before, but for the time
discretization we embed the Euler step described above into a standard
second order Runge-Kutta scheme \cite{shu88}. The stability restriction of this
fully explicit scheme is so that $\Delta t<\alpha\min\left(\Delta x,\Delta y\right)\sqrt{\kappa/\rho_{0}}$
, for some small constant $\alpha$. From this approach we expect
second order global convergence.

\begin{figure}

(a)\epsfig{file=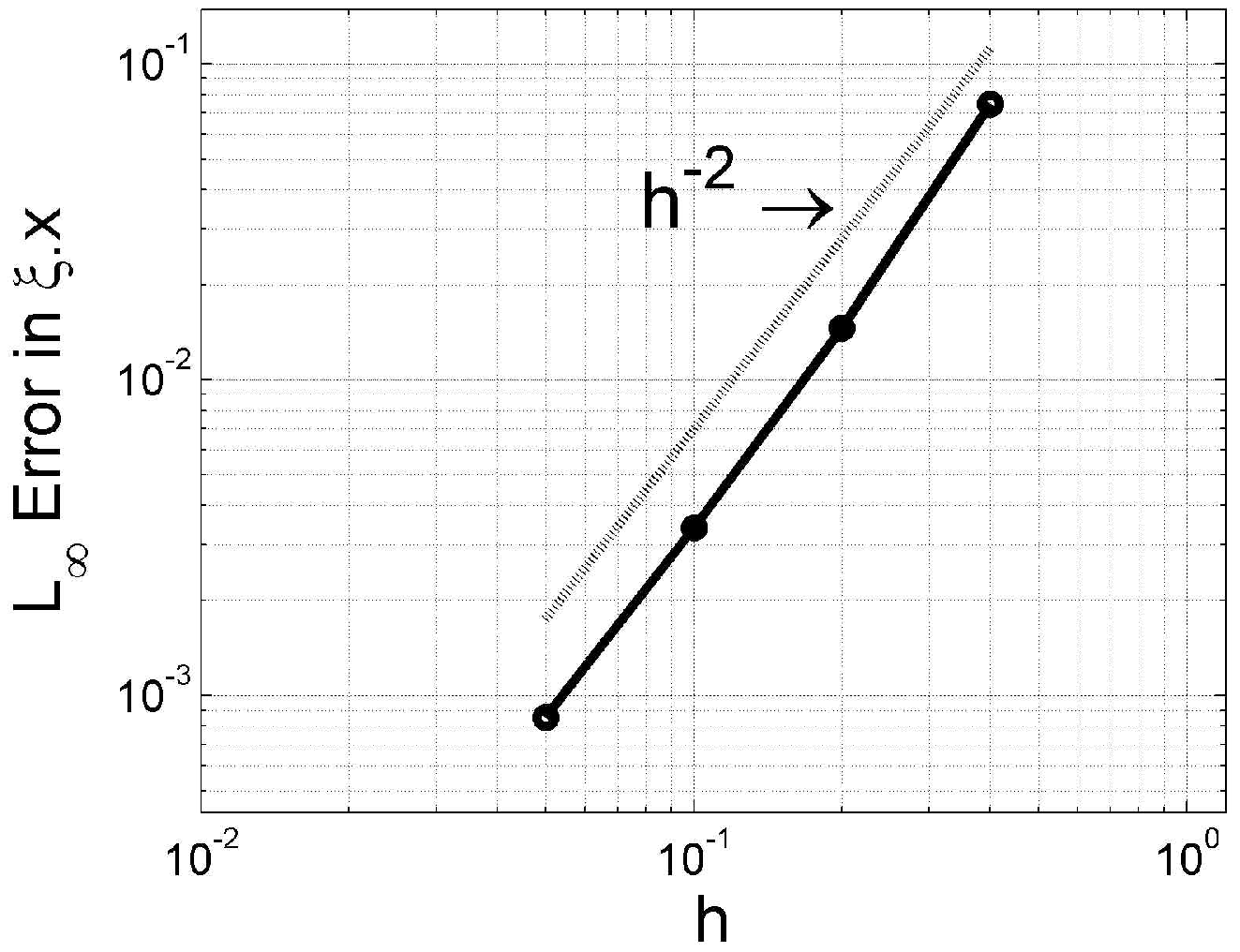, width=2.8in} \ \  (c) \epsfig{file=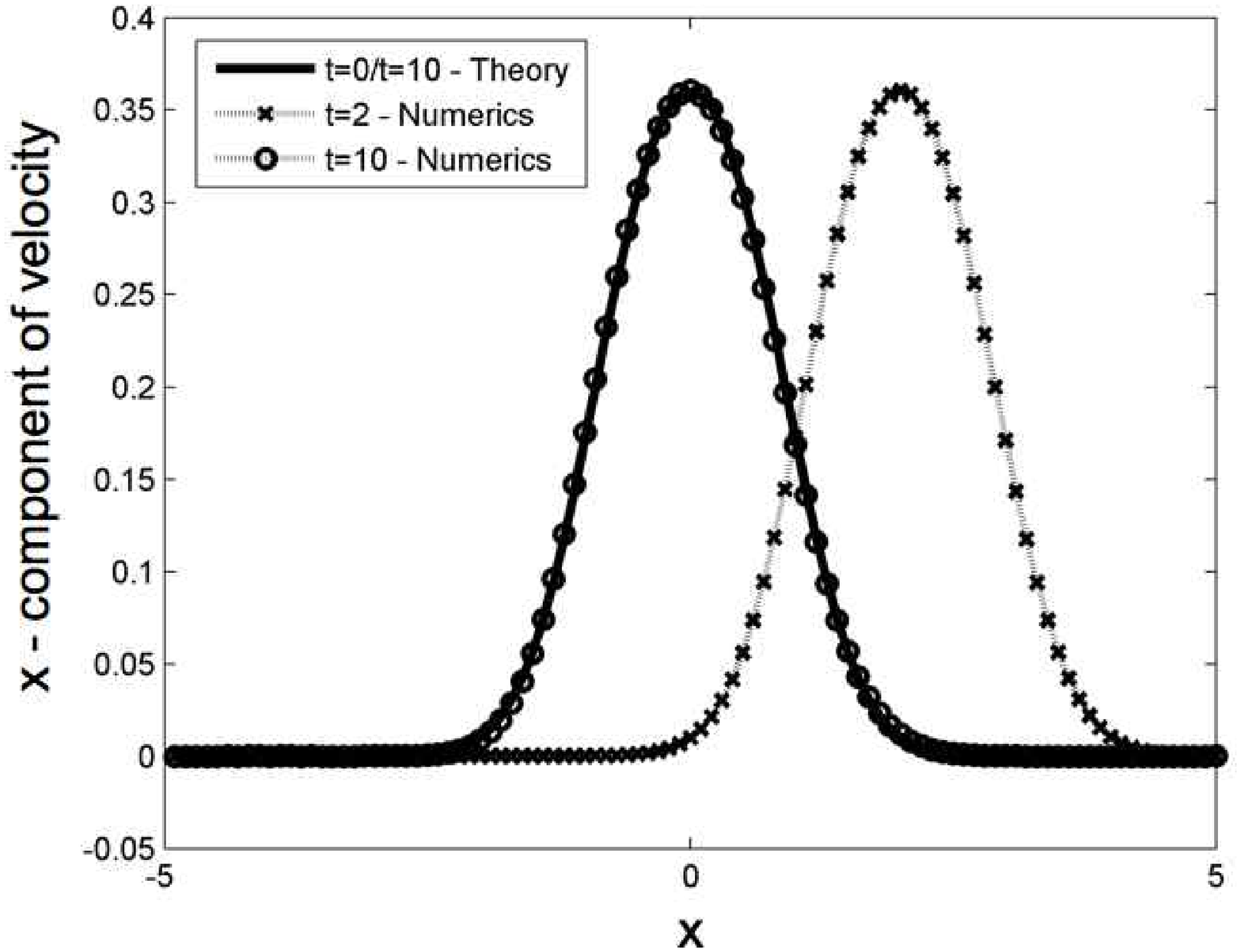, width=2.6in}
\caption{(a) Global convergence of the $L_{\infty}$ norm of the error
  in $\xib$ (connected dots).  We observe a second a order rate of
  convergence.  (b) Travelling $\hat{\vec{x}}\cdot\vec{v}$ wave
  through one full period $t=[0,10]$. Comparison between analytical
  and numerical solutions.}

\label{dynamic}
\end{figure}

In order to verify the convergence rate, we set up a two-dimensional
doubly-periodic domain
$\Omega=\left[-5,5\right]\times\left[-5,5\right]$.  We use Eqs
\ref{eq:xi_exact_solution} and \ref{eq:v_exact_solution} at $t=0$ as
the initial condition, with $c=\kappa=\rho_{0}=1$. The travelling wave
solution should come back to it original shape and location at
$t=10$. For a sequence of grids with $h=\Delta x=\Delta y=\left\{
  \tfrac{2}{40},\tfrac{2}{20},\tfrac{2}{10},\tfrac{2}{5}\right\} $, we
set $\Delta t$ using $\alpha=\frac{1}{10}$, and compute the
$L_{\infty}$error of $\xib$, $E_{\infty}^{\xib}\left(h\right)$ as,

\[
E_{\infty}^{\xib}\left(h\right)=\sup_{\left\{ x,y\right\} \in\Omega}\left|\hat{\vec{x}}\cdot\xib\left(x,y,t=10\right)-\hat{\vec{x}}\cdot\xib\left(x,y,t=0\right)\right|.\]

We report in Figure \ref{dynamic}(a) the expected second order global
convergence. The convergence rate between the two finest grids,
$h=\tfrac{2}{10}$ and $h=\tfrac{2}{5}$, is computed to be $1.97$. We
have also found nearly identical convergence properties for the
velocity; the $x$ velocity's convergence rate is found to be $1.99$
between the two finest grids. We conclude the scheme is globally
second order accurate for all dynamic variables, confirming its
ability to capture dynamic solutions under large-strain elasticity.

Finally, Figure \ref{dynamic}(b) shows one-dimensional cross sections
for $\hat{\vec{x}}\cdot\vec{v}$ at different times.  The solid line
represents the exact solution computed from Eq
\ref{eq:v_exact_solution} at $t=0$, which, by periodicity of the
domain, also corresponds to the solution at $t=10$. We see that the
exact solution and numerical solution agree well for
$h=\tfrac{2}{10}$, a rather coarse grid. We also plot the numerical
solution at $t=2$ for illustrative purposes.

\section{Conclusion and future work}

This work has demonstrated the validity of the reference map for use
in reformulating and simulating solid deformation under a completely
Eulerian framework. There are still several avenues of future
investigation. Other material models are to be simulated, most notably
elasto-plastic laws with and without rate-sensitivity. Also, the
reference map has potential to simplify the simulation of fluid/solid
interactions, due to both phases having a similar Eulerian treatment.
Our preliminary results on this front are promising, and shall be
reported in a future paper. Also, a method to institute traction
boundary conditions within this framework, especially the
traction-free condition, would be important future study.  This may
ultimately be accomplished with a fluid/solid framework, by treating
traction-free boundaries as surfaces of interaction with a
pressure-free, stationary fluid.

\section*{Acknowledgements}

K. Kamrin would like to acknowledge support from the NDSEG and NSF
GRFP fellowship programs. J.-C. Nave would like to acknowledge partial
support by the National Science Foundation under grant DMS-0813648.


\begin{thebibliography}{10}

\bibitem{anand05}
L.~Anand and C.~Su.
\newblock A theory for amorphous viscoplastic materials undergoing finite
  deformations, with application to metallic glasses.
\newblock {\em J. Mech. Phys. Solids}, 53:1362--1396, 2005.

\bibitem{belytschko00}
T.~Belytschko, W.~K. Liu, and B.~Moran.
\newblock {\em Nonlinear Finite Elements for Continua and Structures}.
\newblock J. Wiley and Sons, 2000.

\bibitem{coleman63}
B.~D. Coleman and W.~Noll.
\newblock The thermodynamics of elastic materials with heat conduction and
  viscosity.
\newblock {\em Arch. Ration. Mech. Anal.}, 13:167--178, 1963.

\bibitem{falk98}
M.~L. Falk and J.~S. Langer.
\newblock Dynamics of viscoplastic deformation in amorphous solids.
\newblock {\em Phys. Rev. E}, 57:7192--7205, 1998.

\bibitem{gurtin81}
M.~E. Gurtin.
\newblock {\em An Introduction to Continuum Mechanics}.
\newblock Academic Press, 1981.

\bibitem{haughton93}
D.~M. Haughton.
\newblock {\em Q. Jl Mech. appl. Math}, 46:471--486, 1993.

\bibitem{hoger86}
A.~Hoger.
\newblock The material time derivtaive of logarithmic strain.
\newblock {\em Int. J. Solids Structures}, 22:1019--1032.

\bibitem{koopman08}
A.~J. Koopman, H.~J.~M. Geiselaers, K.~E. Nilsen, and P.~T.~G. Koenis.
\newblock Numerical flow front tracking for aluminium extrusion of a tube and a
  comparison with experiments.
\newblock {\em Int. J. Mater. Form.}, Suppl 1:423--426, 2008.

\bibitem{lee69}
E.~H. Lee.
\newblock Elastic plastic deformation at finite strain.
\newblock {\em J. Appl. Mech.}, 36:1--6, 1969.

\bibitem{lemaitre02}
A.~Lema\^itre.
\newblock Rearrangements and dilatency for sheared dense materials.
\newblock {\em Phys. Rev. Lett.}, 89:195503, 2002.

\bibitem{liu01}
C.~Liu and N.~J. Walkington.
\newblock An eulerian description of fluids containing visco-elastic particles.
\newblock {\em Arch. Rational Mech. Anal.}, 159:229--252, 2001.

\bibitem{liu94}
X.-D. Liu, S.~Osher, and T.~Chan.
\newblock Weighted essentially non-oscillatory schemes.
\newblock {\em J. Comput. Phys.}, 115:200--212, 1994.

\bibitem{meyers06}
A.~Meyers, H.~Xiao, and O.~T. Bruhns.
\newblock Choice of objective rate in single parameter hypoelastic deformation
  cycles.
\newblock {\em Comput. Struct.}, pages 1134--1140, 2006.

\bibitem{miller01}
G.~H. Miller and P.~Colella.
\newblock A high-order eulerian godunov method for elastic–plastic flow in
  solids.
\newblock {\em J. Comput. Phys.}, 167:131--176, 2001.

\bibitem{shu88}
C.-W. Shu and S.~Osher.
\newblock Efficient implementation of essentially non-oscillatory
  shock-capturing schemes.
\newblock {\em J. Comput. Phys.}, 77:439--471, 1988.

\end{thebibliography}
\end{document}